\documentclass[aps,prl,showpacs,twocolumn,amsmath,amssymb]{revtex4}
\usepackage[english]{babel}
\usepackage{latexsym}
\usepackage{epsfig}

\begin{document}

\title{Oscillation regimes of a solid-state ring laser with active beat note stabilization~:\\
from a chaotic device to a ring laser gyroscope}
\author{Sylvain~Schwartz$^1$, Gilles~Feugnet$^1$, Evguenii~Lariontsev$^2$ and Jean-Paul~Pocholle$^1$}

\affiliation{$^1$Thales Research and Technology France, RD 128, F-91767 Palaiseau Cedex, France\\
$^2$Lomonosov State University, Moscow, 119992 Russia}

\date{\today}


\email{sylvain.schwartz@thalesgroup.com}

\begin{abstract}

We report experimental and theoretical study of a rotating diode-pumped Nd-YAG ring laser with active beat note
stabilization. Our experimental setup is described in the usual Maxwell-Bloch formalism. We analytically derive a
stability condition and some frequency response characteristics for the solid-state ring laser gyroscope, illustrating
the important role of mode coupling effects on the dynamics of such a device. Experimental data are presented and
compared with the theory on the basis of realistic laser parameters, showing a very good agreement. Our results
illustrate the duality between the very rich non linear dynamics of the diode-pumped solid-state ring laser (including
chaotic behavior) and the possibility to obtain a very stable beat note, resulting in a potentially new kind of rotation
sensor.

\end{abstract}

\maketitle

\section{Introduction}

Interest in ring lasers developed almost simultaneously with the invention of laser itself \cite{rosenthal, Davis, heer,
Aronowitz_1965, exp_1966}. Intensive work on this device has been motivated both by fundamental aspects (especially in
the field of non linear dynamics, phase transitions, instabilities and chaos \cite{ikeda_1980, hopf, ikeda_1983,
chaos_1985, Lariontsev4, landau_ginzburg, Lar_chaos_1998, topological_phase_1998, Vlad_1998, Mandel_1992}) and by
practical applications (amongst which are the ring laser gyroscope \cite{Aronowitz1970, Aronowitz_2, Chow} and the
single-frequency unidirectional ring laser \cite{clobes, biraben, byer}). The recent achievement of active beat note
stabilization in a diode-pumped Nd-YAG ring laser \cite{schwartz} revived interest for homogeneously broadened (e.g.
solid-state) ring lasers, although this kind of device had already been extensively studied (see for example
\cite{lar_73, Khanin, Khanin_2, Mandel_1988}). In particular, the experiment described in \cite{schwartz} provides both
a simple tool for the study of mode coupling in a resonant macroscopic quantum device such as a toroidal superfluid
\cite{leggett}, and a potentially new kind of ring laser gyroscope involving only standard optical components and no
gaseous medium.

It is well known in the field of homogeneously broadened ring lasers (see references above) that the dynamics of these
devices is mainly ruled by two sources of coupling between the counterpropagating fields, one being due to the
backscattering of light by the cold cavity elements and the other being due to the spatially non uniform saturation of
the gain (or `population inversion grating'). It is also well known that the latter coupling tends to destabilize
bidirectional emission \cite{Siegman}, thus preventing beat note existence and rotation sensing. Although it is not
possible, for a solid-state ring laser, to suppress this coupling in the same way as in the case of a gas ring laser
gyroscope (because of the absence of Doppler gain-broadening), it has been demonstrated in \cite{schwartz}, following
the pioneer work of \cite{lar_83}, that it was however possible to circumvent it by using an additional stabilizing
coupling.

The aim of this paper is to show how fine control of these mode coupling effects can turn the diode-pumped solid-state
ring laser, which has intrinsically a very rich and non linear dynamics, into a stable ring laser gyroscope.

The semi-classical model we use for the description of our device, including active beat note stabilization, is quickly
described in Sec.~II. We then present, in Sec.~III, an experimental overview of the oscillation regimes of the
diode-pumped Nd-YAG ring laser. We show in particular that our data are in good agreement both with previous
experimental observations using lamp-pumped solid-state ring lasers and with theoretical predictions from the
literature. In Sec.~IV, we study both theoretically and experimentally the possibility of stabilizing the beat note.
Sec.~V deals with the frequency response of the solid-state ring laser gyroscope obtained when the stability condition
derived in Sec.~IV is fulfilled. We finally conclude the article in Sec.~VI.

\section{Semi-classical model}

The dynamics of the rotating solid-state ring laser, including the additional stabilizing coupling, can be
satisfactorily described using the typical semi-classical approach \cite{Lamb}.\\

For the laser electric field inside the ring cavity, obeying Maxwell equations, we make the plane wave, uniform field
and slowly-varying envelope approximations. In particular, all the transverse effects and also the longitudinal (i.e.
axial) effects due to the spatial distribution of the laser components will be neglected. We furthermore assume one
single identical longitudinal mode in each direction (this approximation is not valid when the laser is at rest, as will
be discussed in next section) and the same (linear) polarization state \textbf{e}, resulting in the following expression
for the electric field $\mathbf{E}$~:
\begin{equation} \nonumber
\mathbf{E}(x,t) = \textrm{Re} \left[ \tilde{E}_1 (t) e^{i(\omega_c t-kx)} + \tilde{E}_2 (t) e^{i(\omega_c t + kx)}
\right] \mathbf{e} \;,
\end{equation}
where $k=2\pi/\lambda$ is the mean spatial frequency associated with the longitudinal coordinate $x$ and $\omega_c$ is
the mean angular frequency of the emitted modes.

The laser cavity is described, in the framework of Maxwell theory, by a polarization $\mathbf{P}$ due to the active
medium, a dielectric constant $\varepsilon$ and a fictitious conductivity $\kappa$, those parameters being related to
the total cavity loss per time unit $\gamma$ through the relation $\gamma = \kappa / \varepsilon$ and to the frequency
$\omega_c$ through the relation $\omega_c = k/\sqrt{\mu_0\varepsilon}$, where $\mu_0$ is the magnetic permeability of
vacuum. In accordance with the uniform field approximation, the quantities $\kappa$ and $\varepsilon$ are supposed to be
independent of the longitudinal coordinate $x$. However, their possible modulation at the spatial frequency $2k$,
although being usually very small, has to be taken into account for a correct description of the coupling induced by the
cold cavity elements \cite{projection}. In order to avoid unnecessary complexity, we will use the same notation for the
local and mean values of those parameters.

Starting from the typical Maxwell wave equation (using the rationalized MKSA system of units \cite{Jackson})~:
\begin{equation} \label{wave}
\frac{\partial^2\textbf{E}}{\partial t^2} + \frac{\kappa}{\varepsilon} \frac{\partial\textbf{E}} {\partial t}
+\frac{k^2} {\mu_0 \varepsilon}\textbf{E}= - \frac{1}{\varepsilon}\frac{\partial^2 \textbf{P}}{\partial t^2} \;,
\end{equation}
we make a projection on the cavity emission modes, i.e. we multiply equation (\ref{wave}) by $\exp(\pm ikx)$ and
integrate with respect to $x$ along the cavity perimeter. Taking also into account the rotation of the ring laser and
the additional stabilizing coupling, we obtain the following equations of evolution for the slowly-varying amplitudes
$\tilde{E}_1$ and $\tilde{E}_2$~:
\begin{equation} \nonumber
\frac{\textrm{d}\tilde{E}_{1,2}}{\textrm{d}t}=- \frac{\gamma_{1,2}}{2} \tilde{E}_{1,2} + \frac{i\tilde{m}_{1,2}}{2}
\tilde{E}_{2,1} + (-1)^{1,2}\frac{i\Omega}{2} \tilde{E}_{1,2} + \frac{\omega_c \tilde{P}_{1,2}}{2i \varepsilon} \;,
\end{equation}
where $\gamma_1$ and $\gamma_2$ are the counterpropagating modes loss coefficients, and where $\tilde{m}_{1,2}$ are the
cold cavity coupling coefficients, defined as~:
\begin{equation} \label{back}
\tilde{m}_{1,2}= -\frac{\omega_c}{\varepsilon L} \oint_0^L \left[ \varepsilon(x) - \frac{i \kappa(x)}{\omega_c} \right]
e^{-2i (-1)^{1,2} k x} \textrm{d}x \; ,
\end{equation}
$L$ being the cavity length. The rotation-induced angular frequency non-reciprocity $\Omega$ is given by the Sagnac
formula $\Omega = 8\pi A \dot{\theta} / (\lambda L_\textrm{op})$, $A$ being the area enclosed by the cavity,
$L_\textrm{op}$ the cavity optical length and $\dot{\theta}$ the rotation speed. We have also introduced the spatial
harmonics of the complex amplitude of the gain medium polarization $\tilde{P}_{1,2}$, defined by~:
\begin{equation} \nonumber
\tilde{P}_{1,2} = \frac{1}{L} \oint_0^L \mathbf{e} \cdot \mathbf{P} e^{-i\left[\omega_c t + (-1)^{1,2} kx \right]}
\textrm{d} x \;.
\end{equation}
It can be seen, on the mode equations, that the laser dynamics is mainly ruled by three different sources of coupling
between the counterpropagating fields.
\begin{itemize}
\item The coupling induced by the cold cavity (i.e. in the absence of gain), represented by the coefficients
$\tilde{m}_{1,2}$~; as can be deduced from expression (\ref{back}), such a coupling can result for example from
localized losses or from a step of refractive index~; it is well-known in the field of gas ring laser gyroscopes that
this coupling is responsible for a frequency synchronization between the counterpropagating modes at low rotation
speeds, resulting in a zone of non-sensitivity usually called the ``dead zone" \cite{statz_locking}~;
\item The coupling induced by the active medium, represented by the coefficients $\tilde{P}_{1,2}$, whose expression as
a function of the electric field will be derived further in this section, in the framework of the dipolar coupling
theory in quantum mechanics~;
\item The additional coupling introduced in order to stabilize the beat note, which consists, for the
counterpropagating modes, of different loss coefficients $\gamma_{1,2}$, whose mean value $\gamma$ is constant and whose
difference is proportional to the difference between the intensities of the counterpropagating modes~:
\begin{equation} \label{coupl}
\gamma_1-\gamma_2 = a K (|\tilde{E}_1|^2-|\tilde{E}_2|^2) \;,
\end{equation}
where $a$ is the saturation parameter (such that $a|\tilde{E}_{1,2}|^2$ is dimensionless, see further) and $K$ is a
constant chosen to be positive, such that the mode with the higher intensity gets the higher loss coefficient~; this
coupling, first suggested by \cite{lar_83}, has been successfully implemented on a diode-pumped Nd-YAG ring laser
\cite{schwartz}.
\end{itemize}

In order to calculate the polarization of the gain medium, we describe the diode-pumped Nd-YAG crystal as system of
two-level atoms. In accordance with the uniform field approximation, we consider the gain medium and the optical pumping
power to be homogeneously distributed. The gain medium is then fully described by a complex coherence term $\rho_{ab}$
and two real population terms $\rho_{aa}$ and $\rho_{bb}$, $a$ and $b$ referring respectively to the lower and the upper
level of the laser transition, whose frequency will be designated as $\omega_{ab}$. Due to the very short relaxation
time of the lower level of the 1.064~$\mu$m emission line of the Nd-YAG, we will moreover assume $\rho_{aa}=0$. In this
formalism, the macroscopic polarization is given by $\textbf{P} = 2 d n_0 \, \Re \textrm{e} (\rho_{ab}) \, \textbf{e}$
where $d$ is a real number characterizing the dipolar coupling and $n_0$ is the atomic density per volume unit. The
temporal evolution of $\rho_{ab}$ and $\rho_{bb}$ is ruled by the Bloch equations with adiabatic elimination of the
polarization term $\rho_{ab}$ (this is made possible because the coherence damping time $T_2$ is much smaller than the
population inversion relaxation rate $T_1$ and than the electric field decay time inside the cavity $1/\gamma$).
Introducing the population inversion density function $N=n_0 \rho_{ab}$, we obtain~:
\begin{equation} \nonumber
n_0 \rho_{ab} =\frac{i N d T_2}{2 \hbar (1+i\delta)} \left[ \tilde{E}_1(t) e^{i(\omega_c t - kx)} + \tilde{E}_2(t)
e^{i(\omega_c t + kx)} \right] \;,
\end{equation}
where $\delta = T_2(\omega_c-\omega_{ab})$ is the cavity detuning, that will be neglected in our analysis since it is
usually much smaller than unity (typically $\delta \lesssim 10^{-2}$). Defining the spatial average $N_0$ and
$2k$-harmonics $N_{1,2}$ of the population inversion density as~:
\begin{equation} \nonumber
N_0=\frac{1}{L}\oint_0^L N \textrm{d}x \quad \textrm{and} \quad N_{1,2}=\frac{1}{L}\oint_0^L N e^{ 2 i (-1)^{2,1} kx}
\textrm{d}x
\end{equation}
leads to the following expression~:
\begin{equation} \nonumber
\tilde{P}_{1,2}=\frac{i a \hbar} {T_1} \, {(N_0 \tilde{E}_{1,2}+N_{1,2} \tilde{E}_{2,1})} \;,
\end{equation}
where $a$ is the saturation parameter \cite{definitions}. The total mean gain is proportional to $N_0$, while
$N_1=N_2^*$ represents the effects of the population inversion grating.\\

The complete self-consistent equations describing the evolution of the counterpropagating modes eventually read~:
\begin{eqnarray}
\frac{\textrm{d}\tilde{E}_{1,2}}{\textrm{d}t} & \! \!=& \! \!
(-1)^{1,2}\left[\frac{aK}{2}(|\tilde{E}_1|^2-|\tilde{E}_2|^2)
+\frac{i\Omega}{2}\tilde{E}_{1,2}\right]  -\frac{\gamma}{2} \tilde{E}_{1,2} \nonumber \\
& &\!\!\!\!\!\!\!\!\!\!\!+\frac{i\tilde{m}_{1,2}}{2}\tilde{E}_{2,1}
 +\frac{\sigma L}{2T} \left( \tilde{E}_{1,2} N_0+ \tilde{E}_{2,1} N_{1,2} \right)
\label{eq1}
\end{eqnarray}
where $\sigma$ is the emission cross section \cite{definitions}. Concerning the evolution of the population inversion
density function $N(x,t)$, it is ruled by the following equation (obtained from Bloch equations in the secular
approximation)~:
\begin{equation}
\frac{\partial N}{\partial t} = W - \frac{N}{T_1} -\frac{a N}{2 T_1} \left| \tilde{E}_{1} e^{-i k x} + \tilde{E}_{2}
e^{i k x} \right|^2 \;, \label{eq2}
\end{equation}
where $W$ is the pumping rate, and where the second and the third terms stand respectively for the spontaneous and the
stimulated emissions. Throughout this paper, we shall assume the following value for the population inversion lifetime~:
$T_1 = 200$~$\mu$s. Equations (\ref{eq1}) and (\ref{eq2}) will be the starting point for the theoretical description of
the dynamics of the solid-state ring laser with active beat note stabilization in the next sections.

\section{An overview of the oscillation regimes of the diode-pumped Nd-YAG ring laser}

Because of its strongly non linear dynamics, the diode-pumped Nd-YAG ring laser exhibits a broad variety of oscillation
regimes. We report in this section an experimental overview of these regimes, and we discuss for each case the agreement
with previously published data (mainly theoretical studies and experiments with lamp-pumped Nd-YAG ring lasers).

\subsection{Estimation of the relevant laser parameters}

The device we used for our experimental investigations is similar to the one described in reference \cite{schwartz}. It
is made of an approximately 30-cm long stable ring cavity containing a 2.5-cm long Nd-YAG rod placed inside a solenoid.
One of the four cavity mirrors is polarizing and a skew-rhombus geometry is used, with a very small non-planarity angle
(typically $10^{-2}$~rad). The Nd-YAG rod is optically pumped by a 808~nm pigtailed laser diode, and laser emission
occurs at 1.064~$\mu$m. In order to create the additional stabilizing coupling, the current inside the solenoid (and
consequently the difference of losses between the counterpropagating modes) is kept proportional to the difference
between the intensities of the counterpropagating modes by an electronic feedback loop, ensuring condition (\ref{coupl}).
The whole device is placed on a turntable.\\

In order to make comparisons between theory and experiment, it is useful to estimate the typical parameters of our
experimental configuration. Those parameters are mainly the mean loss coefficient $\gamma$, the backscattering
coefficients $\tilde{m}_{1,2}$ and the gain of the feedback loop $K$ (plus the relative excess of pumping power above
threshold $\eta$, which is easily deduced from the current inside the pump laser diode).

\begin{figure}
\begin{center}
\includegraphics[scale=0.65]{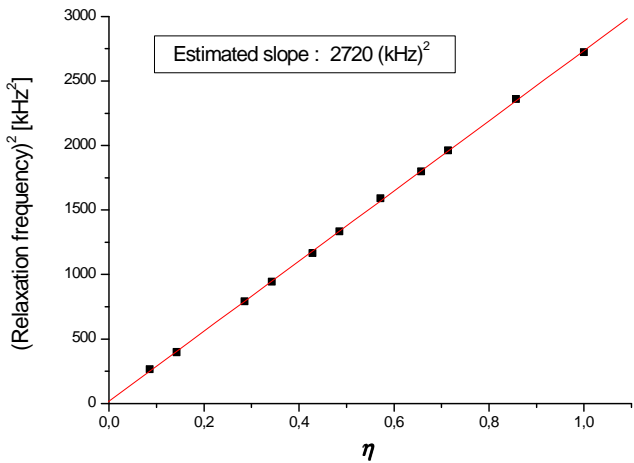}
\end{center}
\caption{(Color online) Square of the relaxation frequency as a function of the pumping rate (data obtained from the
electronic Fourier transform of the laser intensity). The linear dependance is in agreement with
equation~(\ref{formule_relax}), and leads (for $T_1 \simeq 200\mu$s) to the estimated value $\gamma \simeq
21.5$~$10^6$~s$^{-1}$.}\label{relax}
\end{figure}

The loss coefficient can be precisely estimated in a class-B laser thanks to the presence, in the noise spectrum, of
relaxation oscillations at the following frequency~\cite{tang}~:
\begin{equation} \label{formule_relax}
f_r =\frac{\omega_r}{2\pi}= \frac{1}{2\pi} \sqrt{\frac{\gamma \eta}{T_1}}\qquad \textrm{with} \qquad \eta =
\frac{W-W_\textrm{th}}{W_\textrm{th}} \;,
\end{equation}
$W_\textrm{th}$ being the pump power density at laser threshold. The measurement of this frequency as a function of
$\eta$ leads to an estimation of the loss coefficient $\gamma$, as shown on Fig.~\ref{relax}. We obtained the
experimental value $\gamma \simeq 21.5$~$10^6$~s$^{-1}$, which corresponds, for a 30-cm long cavity, to intensity round
trip losses approximately equal to $2.3\%$.

The backscattering coefficient is more difficult to measure than the loss coefficient. However, an estimate can be
obtained by following the same argument as in the case of gas ring lasers \cite{Aronowitz_2}. For this, we first assume,
for symmetry reasons, that $\tilde{m}_1$ and $\tilde{m}_2$ have the same modulus $m$, i.e. we write~:
\begin{equation} \nonumber
\tilde{m}_{1,2} = m e^{i (-1)^{2,1} \theta_{1,2}} \qquad \textrm{with} \qquad m>0 \;.
\end{equation}
Then, we make both hypothesis that the coupling described by $\tilde{m}_{1,2}$ is mainly due to the fraction of light
scattered by the YAG crystal in the solid angle of the couterpropagating beam and that such a scattering is mainly
isotropic. In particular, we neglect the backscattering induced by the cavity mirrors as compared to the backscattering
induced by the crystal. We furthermore assume that all the losses induced in the YAG crystal are due to diffusion, not
absorbtion (this is justified by the fact that the lower level of the laser transition has a very short relaxation time,
typically a few tens of nanoseconds). This leads to the following expression for $m$~:
\begin{equation} \nonumber
m = \frac{c}{L_\textrm{op}} \frac{\lambda \sqrt{b}}{\pi w} \;,
\end{equation}
where $b$ represents the intensity losses corresponding to one pass through the YAG crystal (typically $b = 0.7\%$ for a
25-mm long rod) and $w$ is the waist of the emitted modes (typically 500~$\mu$m). We obtain $m\simeq 5.2$~$10^4$~rad/s,
which corresponds to a few tens of ppm per round trip. Although this method is only a cursory estimate (and cannot
predict in particular any value for $\theta_1$ and $\theta_2$), we will see further in this paper that it provides at
least the correct order of magnitude for $m$.

The strength of the stabilizing coupling $K$ is fully determined by the laser geometry (non-planarity, characteristics
of the solenoid) and by the design of the electronic feedback loop. A good stabilizing effect has been obtained with
$K \simeq 10^7$~s$^{-1}$, which we will use as a reference value in the next sections.\\

We will see further in this paper alternate possibilities to estimate some of the relevant laser parameters (more
precisely $\gamma$, $m$ and $\theta_{1,2}$), based on the study of the laser oscillation regimes.

\subsection{Oscillation regimes in the absence of beat note stabilization}

We first consider the possible oscillation regimes of the ``plain" diode-pumped Nd-YAG ring laser, i.e. in the absence
of additional coupling ($K=0$).\\

\begin{figure}
\begin{center}
\includegraphics[scale=0.415]{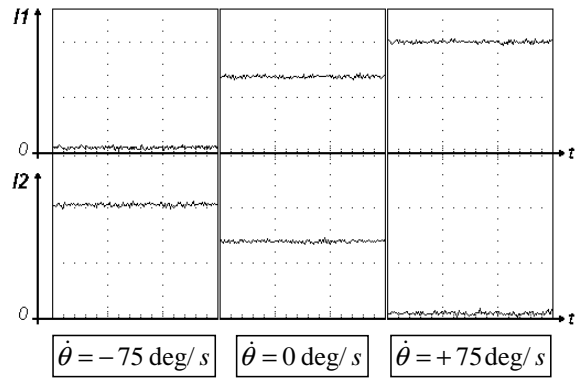}
\end{center}
\caption{Experimental observation of the stationary regimes of the solid-state ring laser. When the laser is at rest
($\dot{\theta}=0$), the bidirectional regime is observed. When the laser is rotating at $\pm 75$~deg/s, unidirectional
operation occurs, the direction of emission depending on the sign of $\dot{\theta}$. The horizontal time scale is
5~$\mu$s/div, while the vertical scale is arbitrary.}\label{stat}
\end{figure}

In the absence of rotation, and when such a laser is operated slightly above threshold ($\eta \simeq 0.2$), we observe a
stable stationary bidirectional regime, as reported on fig.~\ref{stat} (case $\dot{\theta}=0$). The occurrence of such a
regime may seem at first sight surprising, since the analytical condition for the stability of the bidirectional
stationary regime, which reads~\cite{lar_73}~:
\begin{equation} \label{bidir}
m \sin\left| \frac{\theta_1-\theta_2}{2} \right| > \frac{\gamma \eta}{3} \;,
\end{equation}
is obviously not fulfilled with the values of the laser parameters estimated previously. However, it has been shown in
\cite{lar_multi} that the stability condition for the bidirectional stationary regime is weaker when the existence of
many longitudinal modes is accounted. Indeed, experimental measurements with an optical spectrum analyzer showed the
existence of many (typically 3 or 4 per direction) longitudinal modes in the laser at rest, which may explain the fact
that we observe a stable bidirectional emission even if condition (\ref{bidir}) is not fulfilled. It is worth noting
that the laser becomes single-mode in each direction when it is rotated above a critical speed (typically a few deg/s),
ensuring in this case the validity of the theoretical description presented in Sec.~II.
\\

For faster rotations (typically above 70~deg/s), the laser turns to a unidirectional stationary emission regime, which
has been theoretically studied in \cite{Lar_78}. The stability condition for the stationary regime corresponding to
$|\tilde{E}_1|^2\ll |\tilde{E}_2|^2$ reads $\dot{\theta} \sin(\theta_1-\theta_2) < 0 $, while the stability condition
for the opposite case ($|\tilde{E}_1|^2\gg |\tilde{E}_2|^2$) reads $\dot{\theta} \sin(\theta_1-\theta_2) > 0 $. In
particular, the direction of emission depends in this case on the direction of rotation, something which we did observe
experimentally, as shown on fig.~\ref{stat}. Note that references \cite{lar_73} and \cite{Lar_78} predict a ratio of 3
between the intensity of the dominant mode in the unidirectional regime and the intensity of both modes in the
bidirectional regime, while the ratio we measured was only about 1.4. Again, this difference might be explained by the
fact that the single-mode hypothesis, which is used in the theoretical description of references \cite{lar_73} and
\cite{Lar_78}, is not valid when $\dot{\theta} = 0$.\\

\begin{figure}
\begin{center}
\includegraphics[scale=0.47]{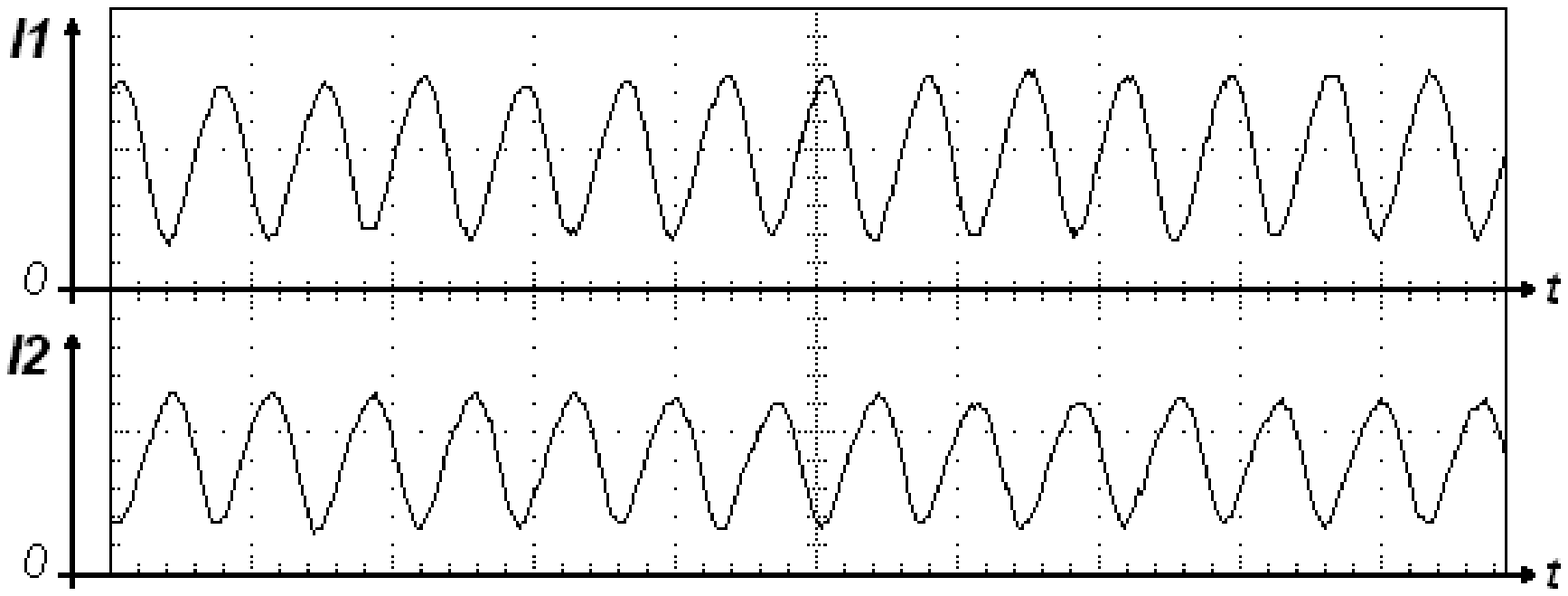}
\end{center}
\caption{Experimental observation of the self-modulation regime of the first kind. The counterpropagating modes
oscillate in phase opposition, with a frequency close to 27~kHz. The horizontal time scale is 50~$\mu$s/div, while the
vertical scale is arbitrary.}\label{selfmod1}
\end{figure}

\begin{figure}
\begin{center}
\includegraphics[scale=0.55]{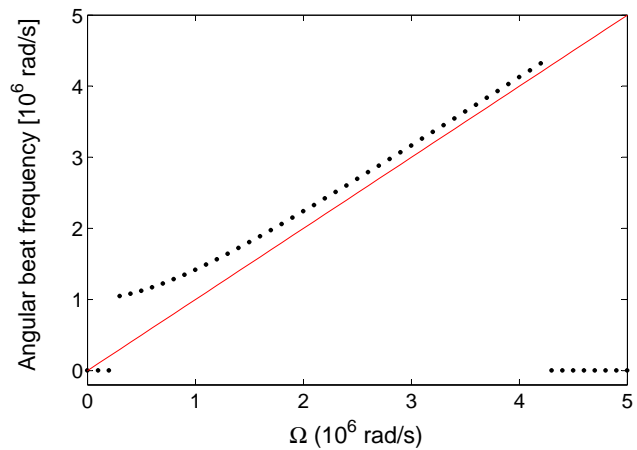}
\end{center}
\caption{(Color online) Numerically computed angular beat frequency as a function of the Sagnac non-reciprocity
$\Omega$, using the following laser parameters~: $\gamma = 2 \; 10^6$~s$^{-1}$, $m=10^6$~rad/s,
$\theta_1-\theta_2=\pi/78$ and $\eta=0.1$. The continuous line corresponds to the ideal Sagnac response. With those
parameters, equation (\ref{omega_lim}) predicts the following values for the boundaries of the beat note zones~:
$\Omega_1 = 0.24\;10^6$~rad/s and $\Omega_2 = 4.2\;10^6$~rad/s, which is in good agreement with this simulation.
Integrating step~: 3~ns. Integrating time~: 7~ms. The plotted values are obtained by averaging the time signals between
5~ms and 7~ms.}\label{3zones}
\end{figure}

In addition to those stationary regimes, a periodic (permanent) regime in which the two counterpropagating modes
oscillate in phase opposition can occur, as reported on fig.~\ref{selfmod1}. This regime, sometimes called
`self-modulation of the first kind', has been described in \cite{lar_81} under the following hypothesis~:
\begin{equation} \label{cond_lim}
|\theta_1-\theta_2|\ll 1 \quad \textrm{,} \quad \gamma\eta\ll m \quad \textrm{and} \quad |\Omega| \ll m  \;.
\end{equation}
It comes out from the theoretical analysis that such a regime can give rise to a beat note when the frequency
non-reciprocity $|\Omega|$ obeys the inequality $\Omega_1 < |\Omega| < \Omega_2$, with~:
\begin{equation} \label{omega_lim}
\Omega_{1,2} = \frac{\gamma' a B}{4 |\theta_1-\theta_2|}+ \frac{(-1)^{2,1} }{2}\sqrt{\left(\frac{\gamma' a B}{2
|\theta_1-\theta_2|}\right)^2-4m^2}
\end{equation}
where $\gamma'=\gamma-m|\theta_1-\theta_2|/2$ and $aB/2 = \eta + 1 - \gamma/ \gamma'$. The numerical simulations we
present on fig.~\ref{3zones} show very good agreement with expression (\ref{omega_lim}). When realistic experimental
parameters are used, numerical simulations show that the laser behavior is qualitatively similar, although conditions
(\ref{cond_lim}) are not fulfilled anymore. In particular, the zone of natural beat note occurrence is still present.
However, we could not observe experimentally a beat note signal that was naturally stable over a reasonably long period,
and to our knowledge such an observation has never been reported in the literature. This is probably due to the weak
stability of this regime over external perturbations (e.g. mechanical noise).
\\

\begin{figure}
\begin{center}
\includegraphics[scale=0.47]{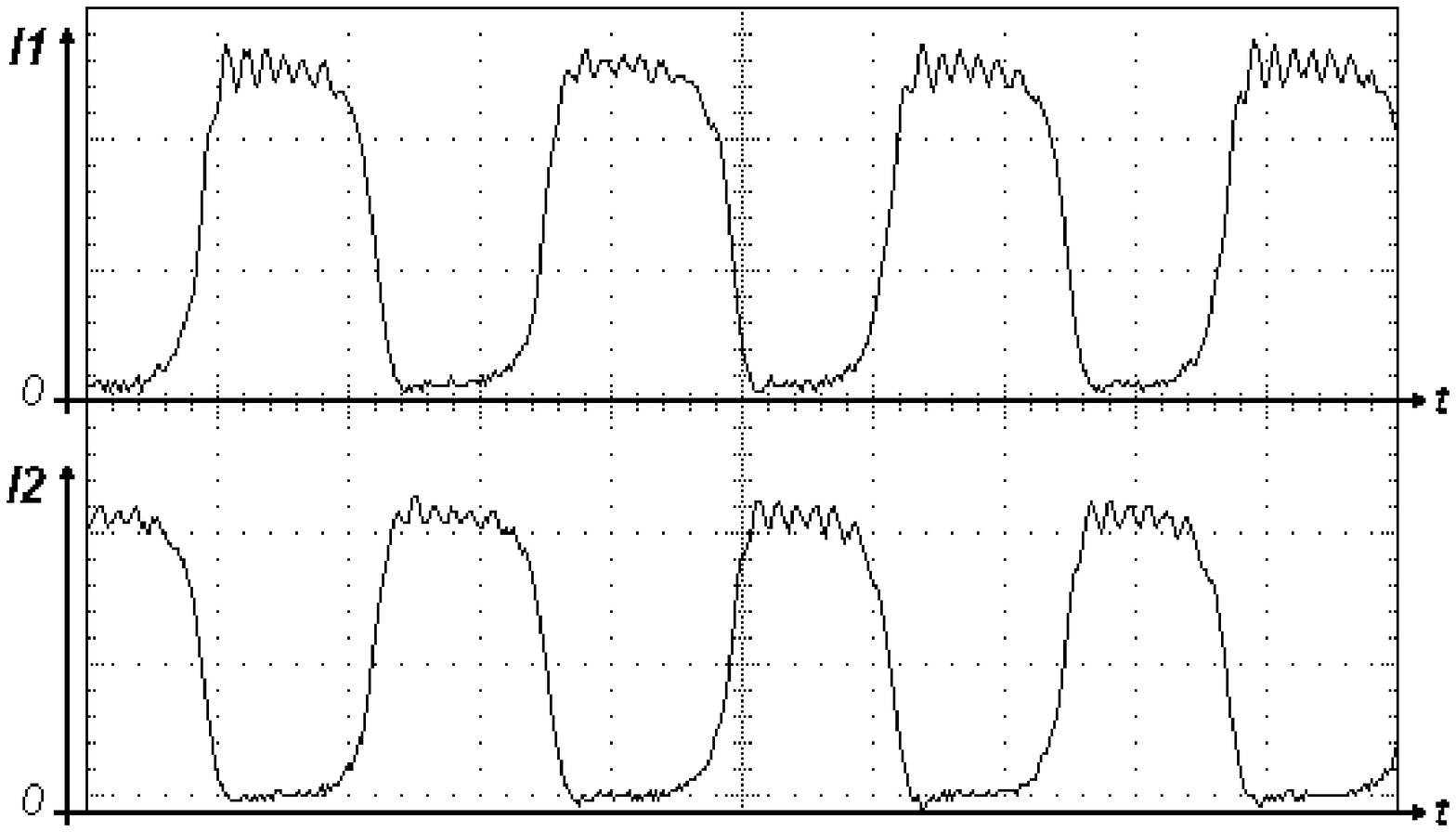}
\end{center}
\caption{Experimental observation of the self-modulation regime of the second kind. The measured switching frequency is
about 3.7~kHz. The horizontal time scale is 100~$\mu$s/div, while the vertical scale is arbitrary.}\label{selfmod2}
\end{figure}

Another periodic regime, typical of solid-state ring lasers, is presented on fig.~\ref{selfmod2}. It consists in a
periodic switch between both unidirectional regimes, with a period approximately equal to $T_1$, and is sometimes called
`self-modulation regime of the second kind'. This regime, which had already been observed on lamp-pumped solid-state
ring lasers \cite{lar_73}, can be described theoretically \cite{Khanin} if one accounts for the existence of a second
line in the emission spectrum of the Nd-YAG (something which is generally neglected invoking thermal equilibrium in the
crystal due to phonon interactions \cite{yag}). Although it is in principle not necessary, we had in practice to
modulate one of the laser parameters (namely the pump power) to observe this regime.
\\

\begin{figure}
\begin{center}
\includegraphics[scale=0.47]{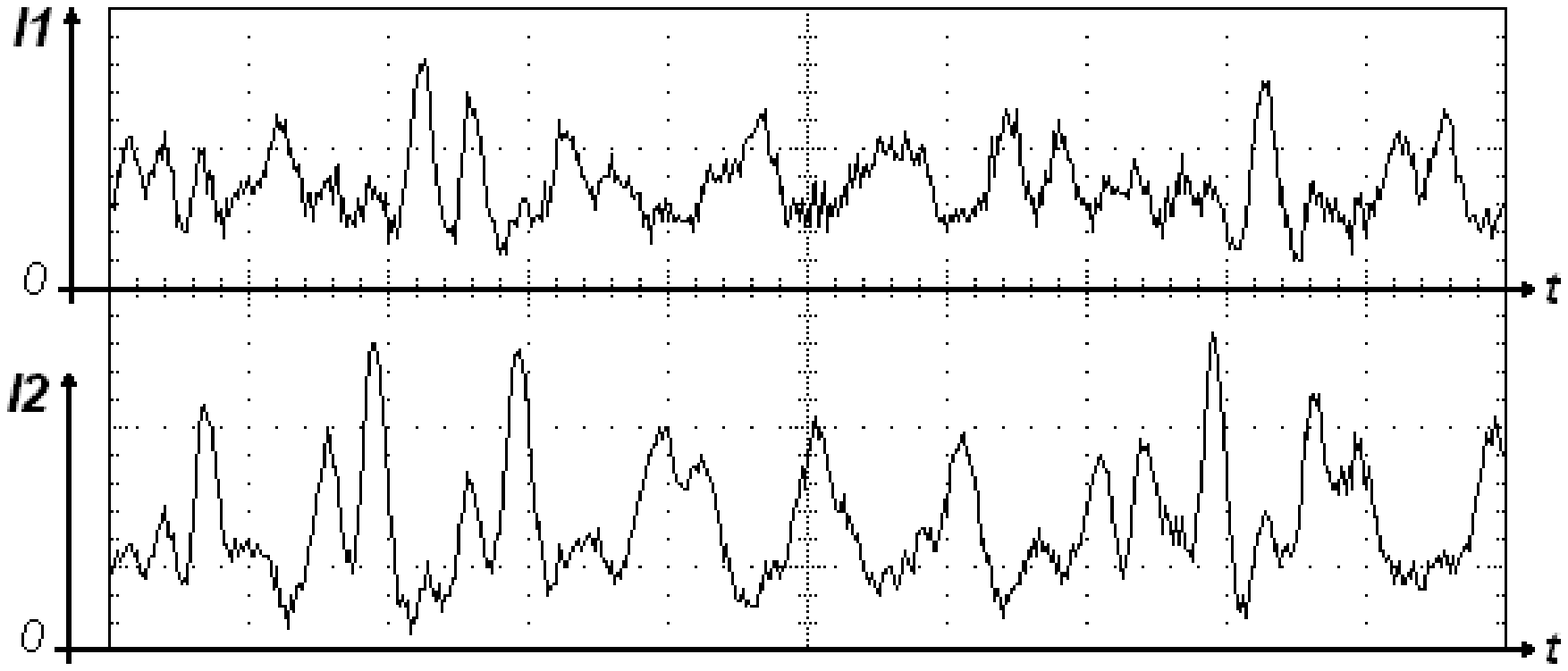}
\end{center}
\caption{Experimental observation of the chaotic behavior of the diode-pumped solid-state ring laser. This regime has
been obtained by periodic modulation of the pump power at a frequency close to the relaxation frequency $f_r$. The
horizontal time scale is 100~$\mu$s/div, while the vertical scale is arbitrary.} \label{chaos}
\end{figure}

Because of the presence of strong non-linearities in the solid-state ring laser, periodic modulation of one of the
parameters can also lead to chaotic behavior, as shown on fig.~\ref{chaos}. Such a regime turns the solid-state ring
laser into a convenient experimental tool for the study of dynamic chaos (see for example reference \cite{chaos_1985}).
Preliminary numerical simulations and experimental work, not reported in this paper, have also shown the possibility of
measuring the Sagnac frequency in the spectrum of the signal obtained by superposing both emitted modes when the laser
is in the chaotic regime. This would present the major advantage of suppressing mode coupling effects, thus improving
the quality of the gyroscopic response. However, some questions about this technique, like for example the problem of
stabilizing the chaotic regime, are still not fully answered at the moment.
\\

\begin{figure}
\begin{center}
\includegraphics[scale=0.65]{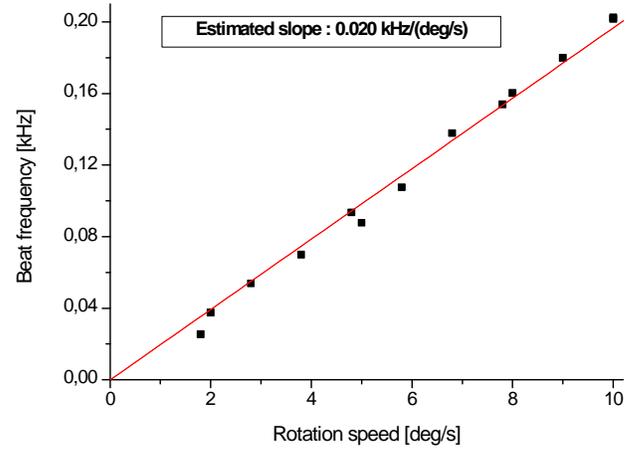}
\end{center}
\caption{(Color online) Measured beat frequency as a function of the rotation rate in the regime of `anomalous scale
factor'. The estimated line slope is about 20~Hz/(deg/s), which is about 100 times smaller than the Sagnac scale factor,
about 2~kHz/(deg/s).}\label{anomalous}
\end{figure}

When the laser is rotating not too fast (typically $|\dot{\theta}|\lesssim 10$~deg/s), we observe a beat note signal,
whose frequency is proportional to the rotation speed but much smaller (about 100 times) than the theoretical Sagnac
frequency (see fig.~\ref{anomalous}). We believe this `anomalous scale factor' regime can be theoretically described if
one accounts for the non-zero relaxation time of the lower level of the laser transition (this time constant, which is
typically on the order of a few tens of nanoseconds, is usually assumed to be equal to zero although it is comparable
with the photon lifetime inside the cavity $1/\gamma \simeq 50$~ns). The establishment of an absorption grating is
likely to give rise to a low frequency beat signal \cite{lar_sat}.
\\

The broad variety of oscillation regimes we observed for the rotating diode-pumped solid-state ring laser illustrates
the richness and the intrinsic non linear characteristics of its dynamics. However, although many of those regimes do
depend on the speed of rotation of the device, none of them exists for a sufficiently broad range of parameters nor is
stable enough to provide a satisfactorily way to use the diode-pumped solid-state ring laser as a rotation sensor. This
observation has been the starting point for our theoretical and experimental work about beat note stabilization, which
will be described in the next sections.

\section{Stabilization of the beat note~: theoretical and experimental study}

We now turn to the study of the diode-pumped solid-state ring laser with active beat note stabilization, i.e. we assume
condition (\ref{coupl}) is fulfilled, with $K>0$. The aim of this section is to show, both theoretically and
experimentally, that under these conditions the beat regime exists and is stable provided the rotation speed is high
enough.

\subsection{Theoretical study}

We use as a starting point for this study equations (\ref{eq1}) and (\ref{eq2}). It is more convenient for analytical
calculation to define new (real) variables as~:
\begin{eqnarray} \nonumber
&Y=|\tilde{E}_1|^2 +|\tilde{E}_2|^2\;, \qquad X=|\tilde{E}_1|^2 -|\tilde{E}_2|^2 \;, & \\
& \Phi = \arg(\tilde{E}_2) - \arg(\tilde{E}_1) \;.&
\end{eqnarray}
We study the beat regime in the limit of high rotation speeds, i.e. we assume~: \begin{equation} \label{cdv} |\Omega|\gg
m \qquad \textrm{and} \qquad |\Omega| \gg \omega_r \;. \end{equation} Under these conditions, the laser parameters in
the beat regime have the following expression~:
\begin{equation} \nonumber
\left\{
\begin{split}
& Y(t) = B(t) + y_M(t) \quad \textrm{with} \quad |y_M| \ll B \;, \\
& X(t) = C(t) + x_M(t) \quad \textrm{with} \quad |C|\, , \,|x_M| \ll B \;, \\
& \Phi (t) - \Omega t = \Phi_0(t) + \Phi_M(t) \quad \textrm{with} \quad |\Phi_0| \, , \, |\Phi_M| \ll 1 \;, \\
\end{split}
\right.
\end{equation}
where the functions $x_M$, $y_M$ and $\Phi_M$ are supposed to oscillate at a frequency close to $\Omega$, while $B$, $C$
and $\Phi_0$ are supposed to be slowly varying with respect to $1/|\Omega|$. The phase origin is chosen such that
$\Phi_0(0)=0$.\\

The spatial harmonics of the population inversion density $N_0$ and $N_1$ can be calculated in this regime, keeping only
the lowest order terms in the expressions for $Y/B$, $X/B$ and $\Phi-\Omega t$ and solving equation (\ref{eq2}). We
obtain~:
\begin{equation} \nonumber
N_0 = WT_1-\frac{aB}{2} N_\textrm{th} \;, \; N_1 = -\frac{aB N_\textrm{th}}{4} \frac{1+i\Omega
T_1}{1+\Omega^2T_1^2}e^{-i\Omega t} \;,
\end{equation}
where $N_\textrm{th} = T_1 W_\textrm{th}$ is the population inversion density at threshold. Integrating equation
(\ref{eq1}) using the same approximation leads to~:
\begin{equation} \label{mod} \left\{
\begin{split}
& y_M = \frac{m}{\Omega}\sqrt{B^2-C^2}\sin\left(\Omega t + \frac{\theta_1 + \theta_2}{2}\right)\sin\left(\frac{\theta_2-
 \theta_1}{2}\right) \\
& x_M = \frac{m}{\Omega}\sqrt{B^2-C^2}\cos\left(\Omega t + \frac{\theta_1 + \theta_2}{2}\right)\cos\left(\frac{\theta_2-
 \theta_1}{2}\right) \\
& \Phi_M = \frac{m}{\Omega}\sqrt{\frac{B-C}{B+C}}\cos\left(\Omega t + \frac{\theta_1 +
\theta_2}{2}\right)\sin\left(\frac{\theta_2- \theta_1}{2}\right) \\
\end{split}
\right.
\end{equation}
Inserting those expressions in equation (\ref{eq1}) up to the first order and then averaging over a few periods of
$1/|\Omega|$, we obtain the following equations for the slowly varying functions $B$ and $C$~:
\begin{equation} \nonumber
\left\{
\begin{split}
\dot{C} & = d C + \frac{B m^2}{2\Omega} \sin (\theta_1- \theta_2) - \frac{a K B C}{2} \;, \\
\dot{B} & = d B + \frac{C m^2}{4\Omega} \sin(\theta_1-\theta_2) - \frac{\sigma l}{4T} \frac{N_\textrm{seuil}
a B^2}{1+\Omega^2T_1^2} - a K C^2 \;, \\
\end{split}
\right.
\end{equation}
where $d$ is defined by $d=\sigma l N_0/T - \gamma$. In the stationary regime, we obtain $a B \simeq 2\eta$, $2d=\gamma
\eta/(1+\Omega^2 T_1^2)$ and~:
\begin{equation} \nonumber
a C = \frac{2 m^2 \sin(\theta_1-\theta_2)}{ \Omega} \left(2K - \frac{\gamma}{1+\Omega^2T_1^2} \right)^{-1} \;.
\end{equation}
The initial hypothesis $|C|\,,\,|x_M|\,,\,|y_M|\ll B$ are self-consistently fulfilled in the high rotation speed limit
defined previously. To study the stability of this solution, we assume a small perturbation $(\delta B,\delta C)\exp(\mu
t)$ and look for the possible values for $\mu$. We find the following three solutions $\mu_{1,2,3}$~:
\begin{equation} \nonumber
\mu_{1,2} = - \frac{1}{2T_1}\pm i \omega_r \quad \textrm{and} \quad \mu_3 = \frac{\gamma \eta / 2 }{1+\Omega^2T_1^2}-
K\eta \;.
\end{equation}
The first two solutions correspond to damped oscillations at the angular frequency $\omega_r$, while the third solution
determines wether or not the beat regime will be stable, the stability condition $\mu_3<0$ reading~:
\begin{equation} \label{cdstab}
2K > \frac{\gamma}{1+\Omega^2T_1^2} \;.
\end{equation}
Physically, this condition expresses the fact that for the beat regime to be stable, the additional stabilizing coupling
(left term) has to be stronger that the destabilizing coupling due to the population inversion grating (right term). It
is a remarkable fact that whatever the stabilizing coupling strength is (provided it is non-zero and positive), the beat
note will always be stable for sufficiently high rotation speeds. This is due to the fact that all the intrinsic
coupling go to zero when the rotation speed increases, while the external stabilizing coupling strength remains constant
whatever the rotation speed is \cite{schwartz}.

\subsection{Experimental achievement}

\begin{figure}
\begin{center}
\includegraphics[scale=0.47]{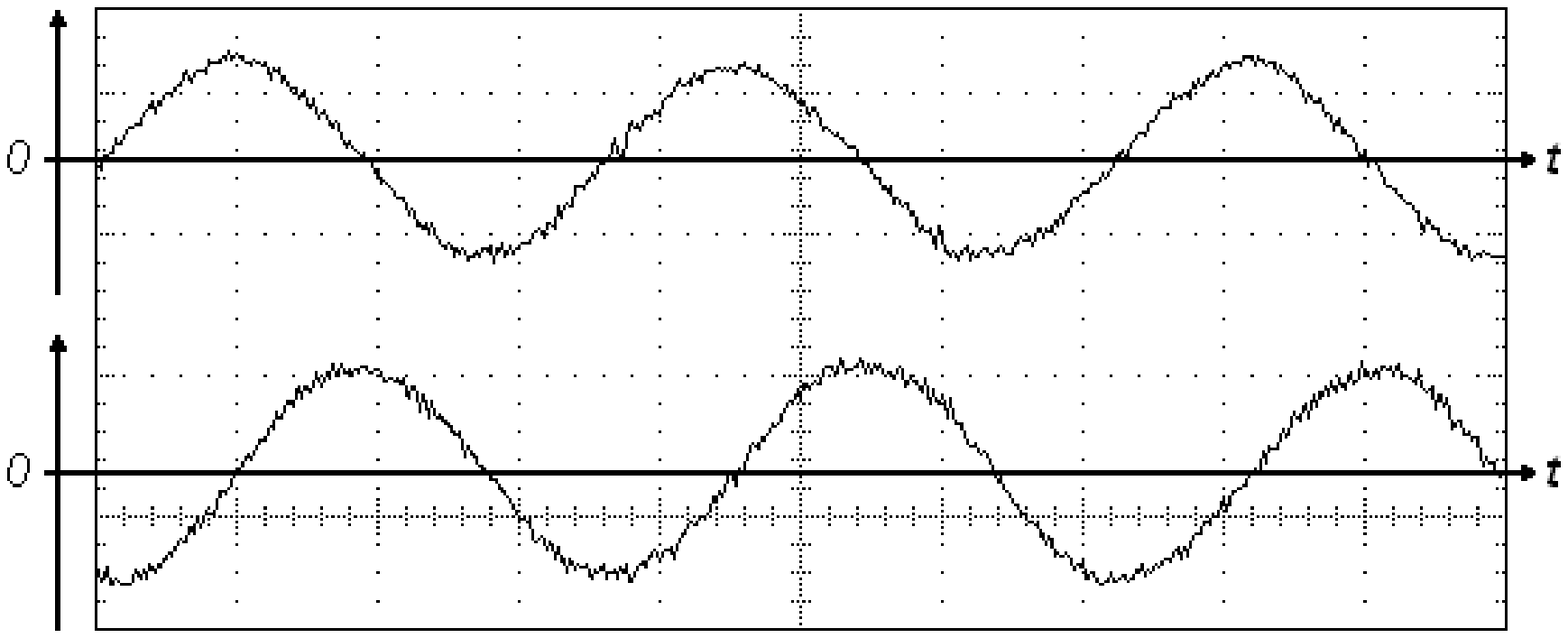}
\end{center}
\caption{Experimental observation of two sinusoidal signals in phase quadrature with two close detectors, which is the
signature of the beat note. The direction of the measured rotation can be deduced from the relative position of the two
signals. Only the AC components of the signals are shown on this figure.} \label{doublebeat}
\end{figure}

When the additional stabilizing coupling is turned on, the beat regime occurs for rotation speeds higher than $\simeq
10$~deg/s. The observation, with two close detectors, of two modulated signals in phase quadrature
(fig.~\ref{doublebeat}) is an undisputable signature of the beat note (as opposed to just intensity modulations), and
gives the additional information of the direction of the rotation.

Experimental study of the intensities of the counterpropagating modes in the beat regime allows to measure the parameter
$\theta_1-\theta_2$. As a matter of fact, it comes out from equations (\ref{mod}) that the relative phase between the
modulated parts of $E_1^2=(X+Y)/2$ and $E_2^2=(Y-X)/2$ is given by $\pi+\theta_1-\theta_2$. For our experimental
configuration, we obtained the following measurement~: $\theta_1-\theta_2 \simeq \pi/20$.

\begin{figure}
\begin{center}
\includegraphics[scale=0.49]{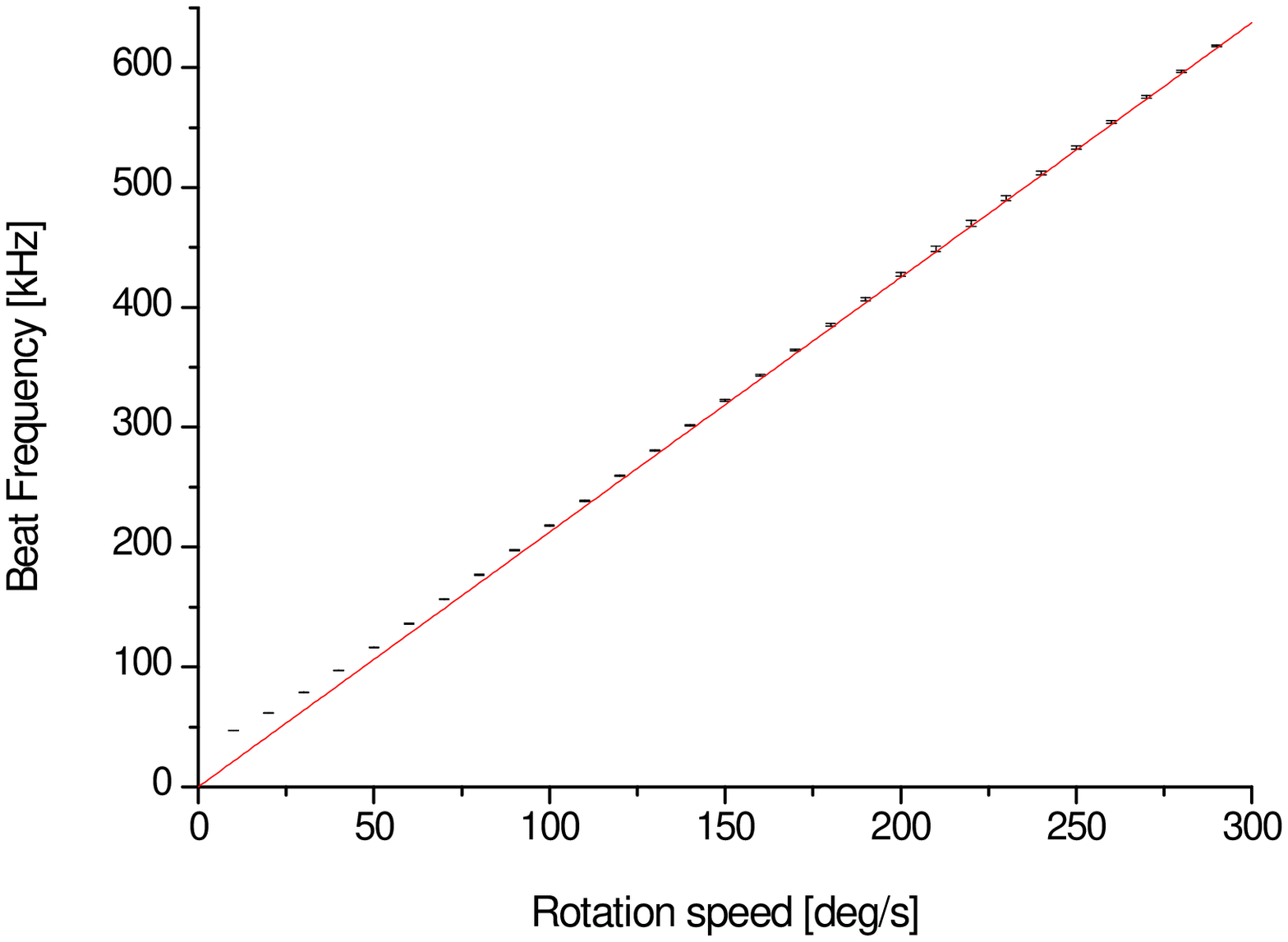}
\end{center}
\caption{(Color online) Experimental frequency response curve of the solid-state ring laser gyroscope, i.e. beat
frequency as a function of rotation speed. The line corresponds to the ideal (theoretical) frequency response
curve.}\label{freqresponse}
\end{figure}

The measured value of the beat frequency as a function of the Sagnac non-reciprocity $\Omega$ is reported on
fig.~\ref{freqresponse}. A very good agreement with the ideal Sagnac line is observed for high values of $|\Omega|$. The
frequency response becomes non-linear when the rotation speed decreases, and finally disappears when $|\dot{\theta}|
\lesssim 10$~deg/s.

It is worth noting that the zone corresponding to the absence of beat note around $\Omega=0$ is not a lock-in zone as in
the case of gas ring laser gyroscopes, but rather a zone of self-modulation of the first kind with an average value of
the phase difference equal to zero. It is also worth noting that the condition for the occurrence of the natural beat
note regime, as described in the previous section, is with our parameters weaker than the validity condition
(\ref{cdv}). The size of the zone of insensitivity to rotation is thus determined by the condition of occurrence for the
natural beat regime, rather than by condition (\ref{cdstab}).

\section{Frequency response of the diode-pumped Nd-YAG ring laser gyroscope}

A typical frequency response curve for the solid-state ring laser is shown on fig.~\ref{freqresponse}. The difference
between the beat frequency and the ideal Sagnac frequency can be expressed analytically provided it is much smaller than
the absolute value of the beat frequency. For this, we derive from equation (\ref{eq1}) the following equation for
$\Phi$~:
\begin{equation} \nonumber
\begin{split}
&\dot{\Phi} = \Omega - \frac{Y}{\sqrt{Y^2-X^2}} \frac{\sigma L}{T} \Im
\textrm{m}(N_1 e^{i\Phi}) \\
&  - \frac{m}{2} \left( \sqrt{\frac{Y-X}{Y+X}}\cos(\Phi + \theta_1) - \sqrt{\frac{Y+X}{Y-X}} \cos(\Phi + \theta_2)
\right) \;.
\end{split}
\end{equation}
Using for the beat regime the definition given previously, we obtain, in the limit $|x_M| \ll B$, $|y_M| \ll B$, $|C|
\ll B$ and $|\Phi_M|\ll 1$, the following equation for $\Phi_0$ (after averaging over a few periods of $1/|\Omega|$)~:
\begin{equation} \nonumber
\dot{\Phi}_0 = \frac{m^2 \cos (\theta_1-\theta_2)}{2 \Omega} + \frac{\gamma \eta}{2 \Omega T_1} \;.
\end{equation}
The beat frequency $\langle \dot{\Phi}\rangle$ is thus finally given by~:
\begin{equation} \label{freq_deviation}
\langle \dot{\Phi} \rangle =\Omega + \frac{m^2 \cos (\theta_1-\theta_2)}{2 \Omega} + \frac{\omega_r^2}{2 \Omega} \;.
\end{equation}
As can be seen on this equation, the two sources of deviation from the ideal Sagnac line are the coupling through
backscattering on the cold cavity elements and the coupling induced by the population inversion grating.

\begin{figure}
\begin{center}
\includegraphics[scale=0.70]{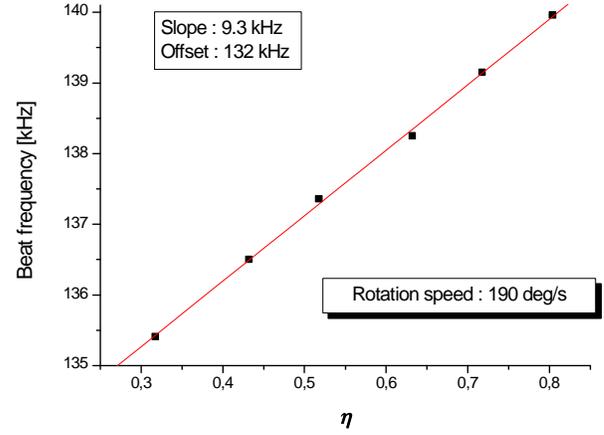}
\end{center}
\caption{(Color online) Experimental dependance of the beat frequency of the solid-state ring laser gyroscope on the
pumping rate for a fixed rotation speed ($\dot{\theta} = 190$~deg/s, corresponding to $\Omega = 8.25\;10^5$~rad/s). The
line results from a linear fit of the data.}\label{freqvseta}
\end{figure}

The linear dependance of the beat frequency on the pumping rate $\eta$ has been checked experimentally, as reported on
fig.~\ref{freqvseta}. We deduce from the line slope another measurement for the loss parameter, namely $\gamma \simeq
19.3 \; 10^6$~s$^{-1}$. This value is in good agreement with the measurement performed on the relaxation frequency
($\gamma \simeq 21.5 \; 10^6$~s$^{-1}$, see Fig.~\ref{relax}). The difference can be attributed to the analytical
approximations made in deriving expression (\ref{freq_deviation}).

The value of the extrapolated beat frequency at $\eta=0$ (offset) provides a measurement of the parameter $m$.
Considering the fact that $\cos(\theta_1-\theta_2) \simeq 1$ (see Sec.~IV), we obtain with this technique $m\simeq 11 \;
10^4$~rad/s, which is twice the value estimated in Sec.~III. This difference might be due in particular to the fact that
we have neglected in Sec.~III the effect of diffusion on crystal edges and on cavity mirrors.

\begin{figure}
\begin{center}
\includegraphics[scale=0.6]{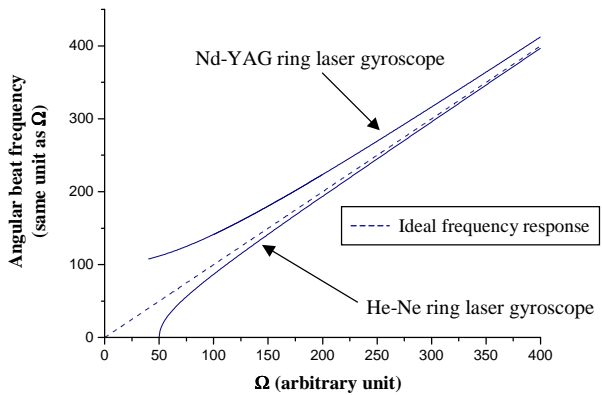}
\end{center}
\caption{(Color online) Graph illustrating the typically admitted shape for the frequency response curve of the gas ring
laser gyroscope, as compared to the frequency response curve of our Nd-YAG ring laser gyroscope. The first curve is
``below" the ideal Sagnac line, while the second is ``above".}\label{freqcomp}
\end{figure}

As can be seen on Fig.~\ref{freqresponse}, the solid-state ring laser has a characteristic response curve ``above" the
ideal Sagnac line, while the typically admitted frequency response curve for the gas ring laser gyroscope is rather
``below" the ideal Sagnac line (Fig.~\ref{freqcomp}). The first reason for this is that the typically admitted picture
for the frequency response curve of the gas ring laser is not always true~: it has been shown \cite{mandel_spreeuw} that
when the coupling induced by the cold cavity elements was dissipative (i.e. when $|\theta_1-\theta_2| \ll 1$), the
frequency response curve was above the ideal Sagnac line, at least in the limit of fast rotations. This result is in
agreement with equation (\ref{freq_deviation}). The second reason is that the deviation induced by the population
inversion grating, which is not present in the gas ring laser gyroscope, is always positive and often dominates the
former coupling, resulting in typical frequency response curves ``above'' the ideal Sagnac line.

\section{Conclusion}

We have studied in this paper the dynamics of a diode-pumped solid-state ring laser with active beat note stabilization.
We have shown in particular that amongst a broad variety of oscillation regimes, including chaotic behavior, a stable
rotation sensitive regime can occur thanks to fine mode coupling control. Focusing on this regime, we have derived
theoretically its stability condition and its frequency response under rotation. We have also studied this regime
experimentally, showing a very good quantitative agreement with our theoretical predictions.

Applications for this work range from the study of non-linear systems with periodic boundary conditions (including other
fields than optics, e.g. superfluidity \cite{schwartz}) to optical gyroscopes. In the latter case, the performances will
depend on the possibility of reducing the strength of the coupling induced by backscattering and by the population
inversion grating. Some techniques to improve these performances will be presented in future publications.

The authors thank Thales Aerospace for constant support. They are also very grateful to Alain Aspect and Philippe Bouyer
for fruitful discussions, and to Mich\`ele Leduc and the French-Russian GDRE Lasers and Optoelectronics for their
contribution to the support of the visits of one of them (E.L.) in France. They also thank François Gutty, Francis
Grabisch and Maxence Rebut for their help in numerical simulations and experiments.


\begin{references}

\bibitem{rosenthal} A. Rosenthal, J. Opt. Soc. Am. \textbf{52} (10), 1143 (1962)

\bibitem{Davis} W. Macek and D. Davis, Appl. Phys. Lett. \textbf{2}, No. 3, 67 (1963)

\bibitem{heer} C. V. Heer, Phys. Rev. \textbf{134} (4A), A799-A804 (1964)

\bibitem{Aronowitz_1965} F. Aronowitz, Phys. Rev. \textbf{139} (3A), A635-A646 (1965)

\bibitem{exp_1966} T. J. Hutchings, J. Winocur, R. H. Durrett, E. D. Jacobs and W. L. Zingery,
Phys. Rev. \textbf{152} (1), 467-473 (1966)

\bibitem{ikeda_1980} K. Ikeda, H. Daido and O. Akimoto, Phys. Rev. Lett. \textbf{45} (9), 709 (1980)

\bibitem{hopf} H. M. Gibbs, F. A. Hopf, D. L. Kaplan and R. L. Shoemaker, Phys. Rev. Lett. \textbf{46} (7), 474 (1981)

\bibitem{ikeda_1983} H. Nakatsuka, S. Asaka, H. Itoh, K. Ikeda and M. Matsuoka, Phys. Rev. Lett. \textbf{50} (2), 109
(1983)

\bibitem{chaos_1985} W. Klische and C.O. Weiss, Phys. Rev. A \textbf{31} (6), 4049 (1985)

\bibitem{Lariontsev4} N. Kravtsov, E. Lariontsev and A. Shelaev, Laser Phys. \textbf{3}, No. 1, 21 (1993)

\bibitem{landau_ginzburg} T.W. Carr and T. Erneux, Phys. Rev. A \textbf{50} (5), 4219 - 4227 (1994)

\bibitem{Lar_chaos_1998} E. Lariontsev, Optics Express \textbf{2} (5), 198 (1998)

\bibitem{topological_phase_1998} V. Toronov and V. Derbov, J. Opt. Soc. Am. B \textbf{15} (4), 1282 (1998)

\bibitem{Vlad_1998} A. Vladimirov, Opt. Comm. \textbf{149}, 67-72 (1998)

\bibitem{Mandel_1992} C. Etrich, P. Mandel, N. Abraham and H. Zeghlache, IEEE Journal of Quantum Electronics
\textbf{28} (4), 811 (1992)

\bibitem{Aronowitz1970} F. Aronowitz and R. Collins, J. Appl. Phys. \textbf{41} (1), 130 (1970)

\bibitem{Aronowitz_2} F. Aronowitz in \textit{Laser applications}, Academic Press, 133 (1971)

\bibitem{Chow} W. Chow, J. Gea-Banacloche, L. Pedrotti, V. Sanders, W. Schleich and M. Scully, Rev. Mod. Phys.
\textbf{57} (1), 61 (1985)

\bibitem{clobes} A. Clobes and M. Brienza, Appl. Phys. Lett. \textbf{21} (6), 265 (1972)

\bibitem{biraben} F. Biraben, Opt. Comm. \textbf{29} (3), 353 (1979)

\bibitem{byer} T. Kane and R. Byer, Opt. Lett. \textbf{10} (2), 65 (1985)

\bibitem{schwartz} S. Schwartz, G. Feugnet, P. Bouyer, E. Lariontsev, A. Aspect and J.-P. Pocholle, Phys. Rev. Lett.
\textbf{97}, 093902 (2006)

\bibitem{lar_73} E. Klochan, L. Kornienko, N. Kravtsov, E. Lariontsev and A. Shelaev, Sov. Phys. JETP \textbf{38} (4),
669 (1974)

\bibitem{Khanin} P. Khandokhin and Y. Khanin, J. Opt. Soc. Am. B \textbf{2}, No. 1, 226 (1985)

\bibitem{Khanin_2} Y. Khanin, J. Opt. Soc. Am. B \textbf{5}, No. 5, 889 (1988)

\bibitem{Mandel_1988} H. Zeghlache, P. Mandel, N.B. Abraham, L.M. Hoffer, G.L. Lippi and T. Mello, Phys. Rev. A \textbf{37}
(2), 470 (1988)

\bibitem{leggett} A. Leggett, Rev. Mod. Phys. \textbf{73}, 307 (2001)

\bibitem{Siegman} A. Siegman, \textit{Lasers}, University Science Books, Mill Valley, California (1986)

\bibitem{lar_83} A. Dotsenko and E. Lariontsev, Sov. J. Quantum Electron. \textbf{14} (1), 117 (1984)

\bibitem{Lamb} W. Lamb, Phys. Rev. \textbf{134}, A1429-A1450 (1964)

\bibitem{projection} R.J.C. Spreeuw, R.C. Neelen, N.J. van Druten, E.R. Eliel and J.P. Woerdman, Phys. Rev. A \textbf{42}
No. 7, 4315 (1990)

\bibitem{Jackson} J. Jackson, \textit{Classical Electrodynamics}, John Wiley and Sons (1975)

\bibitem{Sagnacarticle} G. Sagnac, C.R. Acad. Sci. \textbf{157}, 708 (1913)

\bibitem{statz_locking} H. Haus, H. Statz and W. Smith, IEEE Journal of Quantum Electronics \textbf{21}, No. 1, 78 (1985)

\bibitem{tang} C. Tang, J. Appl. Phys. \textbf{34} (10), 2935 (1963)

\bibitem{lar_multi} E. Klochan, L. Kornienko, N. Kravtsov, E. Lariontsev and A. Shelaev, Radiotekhnika i Elektronika
\textbf{19}, 2096 (1974)

\bibitem{Lar_78} A. Dotsenko, E. Klochan, E. Lariontsev and O. Fedorovich, Radiophysics and Quantum Electronics
\textbf{21} (8), 792 (1979)

\bibitem{lar_81} A. Dotsenko and E. Lariontsev, Sov. J. Quantum Electron. \textbf{11} (7), 907 (1981)

\bibitem{yag} K. Fuhrmann, N. Hodgson, F. Hollinger and H. Weber, J. Appl. Phys. \textbf{62} (10), 4041 (1987)

\bibitem{definitions} The saturation parameter $a$ is defined by $a=T_1T_2d^2\hbar^{-2}/(1+\delta^2)$, while the
emission cross-section $\sigma$ reads $\sigma = a \mu_0 L c \hbar \omega / (L_\textrm{op} T_1)$

\bibitem{lar_sat} E. Klochan, E. Lariontsev, O. Nanii and A. Shelaev, Sov. J. Quantum Electron. \textbf{17} (7), 877
(1987)

\bibitem{mandel_spreeuw} C. Etrich, P. Mandel, R. Centeno Neelen, R.J.C. Spreeuw and J.P. Woerdman, Phys. Rev. A. \textbf{46}
(1), 525 (1992)

\end{references}
\end{document}